\documentclass[%
 reprint,
 amsmath,amssymb,amsfonts
 aps,
pra,
floatfix,
]{revtex4-2}

\usepackage{color}
\usepackage{nicefrac}
\usepackage{graphicx}% Include figure files
\usepackage{dcolumn}% Align table columns on decimal point
\usepackage{bm}% bold math
\usepackage{verbatim}
\usepackage{multirow}
\newcommand{\Langle}{\left\langle}
\newcommand{\Rangle}{\right\rangle}
\newcommand{\eq}[1]{Eq.~(\ref{#1})}
\newcommand{\eqs}[2]{Eqs.~(\ref{#1}) and~(\ref{#2})}

\newcommand{\uvecm}[1]{\hat{\boldsymbol{#1}}}
\newcommand{\GF}[1]{\Gamma\left({#1}\right)}
\newcommand{\Circ}[1]{\operatorname{circ}\left({#1}\right)}

\begin{document}

\preprint{APS/123-QED}

\title{Closed-form expressions for the centroid-tilt error due to scintillation}% Force line breaks with \\
%\thanks{}%

\author{Milo W. Hyde IV}%
 \email{mhyde@epsilonsystems.com}
 \affiliation{Epsilon C5I, Beavercreek, OH 45431, USA}
 
\author{Eric W. Mitchell}%
 \affiliation{Space and Missile Defense Command, Missile Defeat Directorate, Redstone Arsenal, Alabama, USA}%
 \affiliation{The University of Arizona, Wyant College of Optical Sciences, Tucson, Arizona, USA}

\author{Mark F. Spencer}
\affiliation{Department of Engineering Physics, Air Force Institute of Technology, Dayton, OH 45433, USA}

\date{\today}% It is always \today, today,
             %  but any date may be explicitly specified

\begin{abstract}
In adaptive optics, the tracker and wavefront sensor commonly measure irradiance centroids (in their respective focal planes) to estimate the turbulence-degraded wavefront in the pupil plane.  Several factors affect the accuracy of these centroid measurements, including noise, speckle, and scintillation.  The centroid-tilt or ``C-tilt'' errors due to these factors have been studied by numerous researchers; however, to our knowledge, closed-form expressions for the C-tilt error due to scintillation have not been found. 

In this paper, we derive such expressions, assuming spherical-wave illumination of the pupil, a path-invariant index of refraction structure constant $C_n^2$, and Kolmogorov turbulence.  We compare the analytically predicted C-tilt values to those obtained by wave-optics simulations. The agreement, at large, is quite good over a wide range of conditions. As a result, researchers and engineers will find this analysis useful when quantifying the performance of centroid-based trackers and wavefront sensors, both of which are critical adaptive-optics components.
\end{abstract}

%\keywords{Suggested keywords}%Use showkeys class option if keyword
                              %display desired
\maketitle

%\tableofcontents

\section{Introduction}\label{sec:intro}
It is common knowledge that as light propagates through atmospheric turbulence it accumulates phase errors distributed along the path~\cite{Tatarskii:61,1451964,1456150,Ishimaru:99,Roggemann1996,Andrews:05,Sasiela:07}.  Adaptive-optics systems often employ tilt and higher-order wavefront sensors (in the form of trackers and Shack--Hartmann wavefront sensors) to estimate these path-integrated phase errors~\cite{doi:10.2514/1.J061766,Lukin,Roddier,PAO,Hardy,Merritt}. Both sensors generally operate on the same principle: They measure the centroid of an irradiance pattern in a focal plane to estimate the light's phase function in the pupil plane.  For the tracker, the centroid measurement is global and provides an estimate of the average phase gradient over the pupil.  The Shack--Hartmann wavefront sensor, on the other hand, partitions the pupil into subapertures and therefore, provides estimates of the average phase gradients over the subaperture areas.  In the absence of noise, speckle, and scintillation, both of these sensors measure what is known as gradient tilt or ``G-tilt''~\cite{Tyler:94,Sasiela:07}.  

Unfortunately, there are two problems with using G-tilt to estimate the path-integrated phase errors.  For simplicity, we focus the discussion on the tilt sensor or tracker.  The actual or true amount of tilt in the turbulence-degraded wavefront is given by the weights/coefficients of the tip/tilt Zernike polynomials $Z_1^{\pm 1}$~\cite{Noll:76,Sasiela:07, OIAWA,Lakshminarayanan10042011}.  Although G-tilt clearly includes Zernike tilt or ``Z-tilt,'' it also includes contributions from all other off-axis aberrations (predominately coma) that produce shifted or sheered point spread functions~\cite{Herrmann:81,Sasiela:07}.  This Z-tilt, G-tilt error, known as centroid anisoplanatism, has been studied and quantified by many researchers, including Sasiela~\cite{Sasiela:07}, Tyler~\cite{Tyler:94}, and Yura and Tavis~\cite{Yura:85}.      

The second issue with using G-tilt to estimate the path-integrated phase errors (tilt, in particular) concerns the measurement of the irradiance centroid.  Multiple factors affect the accuracy of this measurement, including noise, speckle, and atmospheric amplitude fluctuations (better known as scintillation).  When considering these real-world effects, it is common to distinguish this tilt, called centroid tilt or ``C-tilt,'' from pure G-tilt.  

The error due to C-tilt has also been studied by numerous researchers: Tyler and Fried quantified the C-tilt error due to sensor noise~\cite{Tyler:82}; Burrell \emph{et al.} developed scaling laws for the C-tilt error considering target-induced speckle~\cite{Burrell:23}; and lastly, Tavis and Yura~\cite{Tavis:87}, Holmes~\cite{Holmes:09}, Shaw and Tomlinson~\cite{Shaw:19}, and Mitchell \emph{et al.}~\cite{10.1117/12.3027710,Mitchell:25} studied the effects of amplitude variation on C-tilt.  These latter studies were either numerical in nature (i.e., used quadrature to evaluate integral expressions or wave-optics simulations to quantify C-tilt) or did not consider turbulence-induced scintillation as a potential source of amplitude variation.

In this paper, we derive closed-form expressions that quantify the C-tilt error of a scintillated spherical wave (see Section~\ref{sec:thy}).  To validate these expressions, we employ wave-optics simulations, which we describe in Section~\ref{sec:sim}. We then compare the simulation results to predictions from our analytical expressions in Section~\ref{sec:res}.  The agreement, at large, is excellent over a broad range of conditions. Lastly, we conclude with a brief summary of our work.

\section{Theory}\label{sec:thy}
In this section, we start with Tatarskii's definition of the C-tilt angle~\cite{Tatarskii:61,Tatarskii:71} and utilize Mellin transform techniques~\cite{Sasiela:07,Brychkov2018} to obtain three closed-form expressions for the C-tilt variance, as well as the G-tilt, C-tilt error.  These expressions, in general, are functions of the Fresnel number $N_F$~\cite{IFO,Gaskill1978}.  The first, which is valid for all $N_F$, is in the form of a Meijer G-function~\cite{Sasiela:07,Brychkov2018,Luke1975,Wolfram}, while the second and third are simple, physical relations accurate for $N_F \ll1$ and $N_F \gg 1$, respectively.

\subsection{Centroid-tilt variance}
The angular centroid location in the focal plane is given by
\begin{equation}\label{eq:T0}
\boldsymbol{T}_C = \frac{\displaystyle \iint_{-\infty}^{\infty}\left(\boldsymbol{\rho}/f\right)I\left(\boldsymbol{\rho}\right) \text{d}^2\rho}{\displaystyle \iint_{-\infty}^{\infty}I\left(\boldsymbol{\rho}\right) \text{d}^2\rho},   
\end{equation}
where $\boldsymbol{\rho} = \uvecm{x}x+\uvecm{y}y$, $f$ is the focal length, and $I$ is the irradiance in the focal plane.  Tatarskii showed that \eq{eq:T0} is equivalent to~\cite{Tatarskii:61,Tatarskii:71,Lukin}
\begin{equation}\label{eq:1}
\boldsymbol{T}_C = \frac{1}{k} \dfrac{\displaystyle \iint_{-\infty}^{\infty} \Circ{{{\rho}}/{D}} \nabla \phi\left(\boldsymbol{\rho}\right) I\left(\boldsymbol{\rho}\right) \text{d}^2 \rho}{\displaystyle \iint_{-\infty}^{\infty} \Circ{{{\rho}}/{D}} I\left(\boldsymbol{\rho}\right) \text{d}^2\rho}.
\end{equation}
Here, $\rho$ and $I$ are in the pupil plane, $k = 2 \pi/\lambda$, $D$ is diameter of the circular pupil, and $\nabla \phi$ is the phase gradient.  

Consider a spherical wave radiated by a point source located a distance $z$ away from the pupil.  The irradiance in the pupil plane takes the form
\begin{equation}\label{eq:T2}
    I\left(\boldsymbol{\rho}\right) = \frac{\left|U_0\right|^2}{\left(\lambda z\right)^2} \exp\left[2\chi\left(\boldsymbol{\rho}\right)\right],
\end{equation}
where $U_0$ is the amplitude of the point source and $\chi$ is the log-amplitude fluctuations due to atmospheric turbulence~\cite{Ishimaru:99,Andrews:05}.  Substituting  \eq{eq:T2} into \eq{eq:1} yields
\begin{equation}\label{eq:T4}
\boldsymbol{T}_C = \frac{1}{k} \dfrac{\displaystyle \iint_{-\infty}^{\infty} \Circ{{{\rho}}/{D}} \nabla \phi\left(\boldsymbol{\rho}\right) \exp\left[2\chi\left(\boldsymbol{\rho}\right)\right] \text{d}^2 \rho}{ \displaystyle \iint_{-\infty}^{\infty} \Circ{{{\rho}}/{D}} \exp\left[2\chi\left(\boldsymbol{\rho}\right)\right] \text{d}^2\rho}.    
\end{equation}
We are interested in the C-tilt variance (note that the mean equals zero); therefore,
\begin{widetext}
\begin{equation}\label{eq:2}
\langle \boldsymbol{T}_C \cdot \boldsymbol{T}_C \rangle = \langle T_C^2 \rangle = \frac{1}{k^2} \Langle \dfrac{\displaystyle \iiiint_{-\infty}^{\infty}  \Circ{{{\rho_1}}/{D}}\Circ{{{\rho_2}}/{D}} \nabla_1 \phi\left(\boldsymbol{\rho}_1\right) \cdot \nabla_2 \phi\left(\boldsymbol{\rho}_2\right)  \exp\left[2 \chi\left(\boldsymbol{\rho}_1\right) \right] \exp\left[2 \chi\left(\boldsymbol{\rho}_2\right) \right]  \text{d}^2 \rho_1 \text{d}^2 \rho_2}{\displaystyle \iiiint_{-\infty}^{\infty}  \Circ{{{\rho_1}}/{D}}\Circ{{\rho_2}/{D}} \exp\left[2 \chi\left(\boldsymbol{\rho}_1\right) \right] \exp\left[2 \chi\left(\boldsymbol{\rho}_2\right) \right]  \text{d}^2 \rho_1 \text{d}^2 \rho_2}\Rangle.
\end{equation}
To proceed, we make the approximation that the mean of the quotient is equal to the quotient of the means, such that
\begin{equation}\label{eq:T3}
\langle T_C^2 \rangle \approx \frac{1}{k^2}  \dfrac{\displaystyle \iiiint_{-\infty}^{\infty}  \Circ{{{\rho_1}}/{D}}\Circ{{{\rho_2}}/{D}} B_{\nabla \phi}\left(\left|\boldsymbol{\rho}_1-\boldsymbol{\rho}_2\right|\right)  \exp\left[4 B_\chi\left(\left|\boldsymbol{\rho}_1-\boldsymbol{\rho}_2\right|\right) \right]  \text{d}^2 \rho_1 \text{d}^2 \rho_2}{\displaystyle \iiiint_{-\infty}^{\infty}  \Circ{{{\rho_1}}/{D}}\Circ{{\rho_2}/{D}} \exp\left[4 B_\chi\left(\left|\boldsymbol{\rho}_1-\boldsymbol{\rho}_2\right|\right) \right] \text{d}^2 \rho_1 \text{d}^2 \rho_2},
\end{equation}
\end{widetext}
where $B_{\nabla \phi}$ and $B_{\chi}$ are the phase gradient and log-amplitude covariance functions~\cite{Tyler:94, Fried:67}.  Following Holmes~\cite{Holmes:09}, we also assume that the turbulent phase gradient and log amplitude are statistically independent.  Making the change of variables $\boldsymbol{\rho}'=\boldsymbol{\rho}_1$ and $\boldsymbol{\rho}=\boldsymbol{\rho}_1-\boldsymbol{\rho}_2$ and transforming to polar coordinates produces
\begin{equation}\label{eq:3}
\langle T_C^2 \rangle =\frac{1}{k^2} \frac{\displaystyle \frac{2 \pi}{A} \int_{0}^{\infty} \rho \Lambda\left(\rho/D\right)  B_{\nabla \phi}\left({\rho}\right) \exp\left[ 4 B_{\chi}\left({\rho}\right) \right] \text{d}\rho}{\displaystyle \frac{2 \pi}{A} \int_{0}^{\infty} \rho \Lambda\left(\rho/D\right)  \exp\left[ 4 B_{\chi}\left({\rho}\right) \right] \text{d}\rho},   
\end{equation}
where $A = \pi\left(D/2\right)^2$ is the area of the pupil and 
\begin{equation}\label{eq:4}
\Lambda\left(x\right) = \frac{2}{\pi}\left[\cos^{-1}\left(x\right)-x\sqrt{1-x^2}\right]\operatorname{circ}\left(\frac{x}{2}\right).
\end{equation}
is the optical transfer function~\cite{IFO,Gaskill1978}.  Equation~\eqref{eq:3} is the primary focus of this paper.   

\subsection{Evaluating \eq{eq:3}}
To proceed further, we make the weak turbulence approximation $\exp\left[ 4 B_{\chi}\left({\rho}\right) \right] \approx 1 + 4 B_{\chi}\left({\rho}\right)$, as it is not possible to evaluate the integrals with $B_{\chi}$ in the arguments of the exponentials.  We focus first on the numerator.  

\subsubsection{The numerator}
The weak turbulence approximation splits the numerator of \eq{eq:3} into the sum of two integrals:
\begin{equation}\label{eq:5}
\begin{gathered}
\langle T_C^2 \rangle_\text{N} \approx \frac{2 \pi}{k^2 A} \int_{0}^{\infty} \rho \Lambda\left(\rho/D\right)  B_{\nabla \phi}\left({\rho}\right) \text{d}\rho \hfill \\
\; +\, \frac{2 \pi}{k^2 A} \int_{0}^{\infty} \rho \Lambda\left(\rho/D\right)  B_{\nabla \phi}\left({\rho}\right) 4B_{\chi}\left({\rho}\right) \text{d}\rho. \hfill
\end{gathered}
\end{equation}
We recognize the first integral as the G-tilt variance $\Langle T_G^2\Rangle$~\cite{Tatarskii:71,Tyler:94,Sasiela:07}.  For convenience, we refer to the second as the G-tilt, C-tilt variance $\Langle T_{GC}^2\Rangle$.

Starting with $\Langle T_G^2\Rangle$, we can evaluate the integral rather easily using the spherical-wave version of Tyler's expression for $B_{\nabla \phi}$~\cite{Tyler:94}, i.e.,
\begin{equation}\label{eq:6}
B_{\nabla \phi}\left({\rho}\right) = \frac{5}{24} \sqrt{\pi} \Gamma\begin{bmatrix}1/6\\2/3\end{bmatrix} C_n^2 k^2 z \rho^{-1/3},    
\end{equation}
where the gamma function notation is~\cite{Sasiela:07}
\begin{equation}
\Gamma\begin{bmatrix} a_1,a_2,\cdots,a_m\\b_1,b_2,\cdots,b_n\end{bmatrix} = \frac{\GF{a_1} \GF{a_2} \cdots \GF{a_m}}{\GF{b_1} \GF{b_2} \cdots \GF{b_n}}.
\end{equation}
Although somewhat obvious, $B_{\nabla \phi}$ in \eq{eq:6} assumes Kolmogorov turbulence and path-independent index of refraction structure constant $C_n^2$.  Substituting \eq{eq:6} into $\Langle T_G^2 \Rangle$ and making the change of variables $\rho = Dx$ yields
\begin{equation}
\Langle T_G^2 \Rangle = \frac{5}{3}\sqrt{\pi} \Gamma\begin{bmatrix}1/6\\2/3\end{bmatrix} C_n^2 z D^{-1/3} \int_0^\infty \frac{\text{d}x}{x} x^{5/3} \Lambda\left(x\right).  
\end{equation}
The integral is equivalent to the Mellin transform of $\Lambda\left(x\right)$ evaluated at $5/3$.  Reference~\cite{Sasiela:07} gives this Mellin transform.  Making the requisite substitutions reveals
\begin{equation}\label{eq:9}
   \Langle T_G^2 \Rangle = \Gamma\begin{bmatrix}1/6,4/3\\2/3,17/6\end{bmatrix} C_n^2 z D^{-1/3}.
\end{equation}

Proceeding to the G-tilt, C-tilt variance in \eq{eq:5}, we substitute in $B_{\nabla \phi}$ from \eq{eq:6} and again make the change of variables $\rho = Dx$ producing
\begin{equation}\label{eq:10}
\begin{gathered}
\Langle T_{GC}^2 \Rangle = \frac{5}{3}\sqrt{\pi} \Gamma\begin{bmatrix}1/6\\2/3\end{bmatrix} C_n^2 z D^{-1/3} \hfill \\
\quad \times \int_0^\infty x^{2/3} \Lambda\left(x\right) 4 B_\chi\left(Dx\right) \text{d}x. \hfill
\end{gathered}
\end{equation}
The integral in \eq{eq:10} can be evaluated using the Mellin convolution theorem~\cite{Sasiela:07,Brychkov2018}.  To do so, we need the Mellin transforms of the functions in the integrand,  $\Lambda\left(x\right)$ and $4 B_\chi\left(x\right)$---the former we have, the latter we do not.  Therefore, we need to derive the Mellin transform of $4 B_\chi$ to proceed further.  

%\subsection{Mellin transform of $4B_\chi$}
The (weak-turbulence) spherical-wave log-amplitude covariance is given by~\cite{Fried:67}
\begin{equation}\label{eq:11}
\begin{gathered}
4 B_\chi\left(x\right) = 16 \pi^2 k^2 \int_0^z \int_0^\infty \kappa \Phi_n\left(\kappa,\zeta\right) J_0\left(\frac{\zeta}{z}\kappa x\right) \hfill \\
\quad \times \sin^2\left[\frac{z}{2k} \frac{\zeta}{z}\left(1-\frac{\zeta}{z}\right)\kappa^2\right]\text{d}\kappa \text{d}\zeta. \hfill    
\end{gathered}
\end{equation}
For constant $C_n^2$ and Kolmogorov turbulence, \eq{eq:11} specializes to
\begin{equation}\label{eq:12}
\begin{gathered}
4 B_\chi\left(x\right) = 2^{8/3} \sqrt{\pi} \frac{5}{9} \Gamma\begin{bmatrix}5/6\\2/3\end{bmatrix} C_n^2 k^2 \int_0^z \int_0^\infty \kappa^{-8/3} \hfill \\
\quad \times J_0\left(\frac{\zeta}{z}\kappa x\right)   \sin^2\left[\frac{z}{2k} \frac{\zeta}{z}\left(1-\frac{\zeta}{z}\right)\kappa^2\right]\text{d}\kappa \text{d}\zeta. \hfill
\end{gathered}
\end{equation}

We can use the Mellin convolution theorem to write the integral over $\kappa$ as an equivalent contour integral, such that
\begin{widetext}
\begin{equation}\label{eq:13}
\begin{gathered}
4 B_\chi\left(x\right) = -\frac{5\pi}{72} \Gamma\begin{bmatrix}5/6\\2/3\end{bmatrix} C_n^2 k^2 x^{5/3} \frac{1}{\text{j}2\pi} \int_C x^{-s} \left(\frac{k}{2z}\right)^{-s/2}   
\Gamma\begin{bmatrix}s/2-5/6,-s/4\\-s/2+11/6,s/4+1/2\end{bmatrix} \hfill \\
\quad \times \int_0^z \left(\frac{\zeta}{z}\right)^{5/3-s/2}\left(1-\frac{\zeta}{z}\right)^{s/2}\text{d}\zeta \text{d}s,\hfill
\end{gathered}    
\end{equation}
where the contour $C$ crosses the real $s$ axis between $5/3 < \operatorname{Re}\left(s\right) < 19/6$.  The $\zeta$ integral is equal to a beta function~\cite{Gradshteyn2000}, which can be written as the product of gamma functions, namely,
\begin{equation}\label{eq:14}
\begin{gathered}
\int_{0}^z \left(\frac{\zeta}{z}\right)^{5/3-s/2}\left(1-\frac{\zeta}{z}\right)^{s/2}\text{d}\zeta  = z \Gamma\begin{bmatrix}8/3-s/2,1+s/2\\11/3\end{bmatrix}. 
\end{gathered}        
\end{equation}
Substituting \eq{eq:14} into \eq{eq:13} and performing some minor algebra to write the expression in the form of an inverse Mellin transform~\cite{Sasiela:07,Brychkov2018} yields
\begin{equation}\label{eq:155}
\begin{gathered}
x^{-5/3} 4 B_\chi\left(x\right) = \frac{1}{\text{j}2\pi} \int_C \text{d}s \left\{ -\frac{5\pi}{72} \Gamma\begin{bmatrix}5/6\\2/3,11/3\end{bmatrix} C_n^2 k^2 z  
 \left(\frac{k}{2z}\right)^{-s/2} \Gamma\begin{bmatrix}s/2-5/6,-s/4,-s/2+8/3,s/2+1\\-s/2+11/6,s/4+1/2\end{bmatrix} \right\} x^{-s}. 
\end{gathered}    
\end{equation}
\end{widetext}
The quantity in braces is the Mellin transform of $x^{-5/3} 4 B_\chi\left(x\right)$.  

We can now evaluate $\Langle T_{GC}^2 \Rangle$ in \eq{eq:10}.  First, we multiply and divide $4 B_\chi\left(Dx\right)$ in the integrand by $\left(Dx\right)^{-5/3}$ producing
\begin{equation}
\begin{gathered}
\Langle T_{GC}^2 \Rangle = \frac{5}{3}\sqrt{\pi} \Gamma\begin{bmatrix}1/6\\2/3\end{bmatrix} C_n^2 z D^{4/3} \hfill \\
\quad \times \int_0^\infty x^{7/3} \Lambda\left(x\right)  \left[\left(Dx\right)^{-5/3} 4 B_\chi\left(Dx\right) \right] \text{d}x. \hfill
\end{gathered}
\end{equation}
Then, applying the Mellin convolution theorem and the Mellin transforms for $\Lambda\left(x\right)$ and $x^{-5/3} 4 B_\chi\left(x\right)$, we obtain
\begin{equation}\label{eq:17}
\begin{gathered}
\Langle T_{GC}^2 \Rangle = -2^{-11/2} \frac{25}{27} \Gamma\begin{bmatrix}1/6,5/6\\2/3,2/3,11/3\end{bmatrix} \left(C_n^2 k zD^{2/3}\right)^2 \hfill \\
\; \times \frac{1}{\text{j}2\pi} \int_C \left(N_F^2\right)^{-s} \Gamma\begin{bmatrix}-5/12+s,1/12+s,1+s\end{bmatrix} \hfill \\
\; \times\, \Gamma\begin{bmatrix}-s,5/6-s,13/12-s,4/3-s,19/12-s\\11/12-s,17/12-s,11/6-s,7/3-s\end{bmatrix} \text{d}s, \hfill
\end{gathered}
\end{equation}
where $N_F = \pi \left(D/2\right)^2/\left(\lambda z\right)$ is the Fresnel number~\cite{IFO,Gaskill1978} and $C$ crosses the real $s$ axis between $5/12 < \operatorname{Re}\left(s\right) < 19/24$.  

For relative ease of numerical computation, we can write \eq{eq:17} in the form of a Meijer G-function~\cite{Wolfram,Sasiela:07, Gradshteyn2000,Brychkov2018}.  Numerical routines to evaluate such functions are available in MATLAB, Mathematica, and Python.  Note that there are several different definitions of a Meijer G-function.  Here, we use the one given in Ref.~\cite{Wolfram}, which is consistent with the Meijer G-function routines in Mathematica and Python, namely,
\begin{equation}\label{eq:18}
\begin{gathered}
G^{m,n}_{p,q}\left(z\left|\begin{matrix} a_1, a_2, \cdots, a_p\\b_1, b_2, \cdots, b_q\end{matrix}\right.\right) = \frac{1}{\text{j}2\pi} \int_\gamma \frac{\prod_{j=1}^n\Gamma\left(1-a_j-s\right)}{\prod_{j=n+1}^p\Gamma\left(a_j+s\right)} \hfill \\
\quad \times\, \frac{\prod_{j=1}^m\Gamma\left(b_j+s\right)}{\prod_{j=m+1}^q\Gamma\left(1-b_j-s\right)} z^{-s} \text{d}s. \hfill  
\end{gathered}
\end{equation}

The integration contour $\gamma$ in \eq{eq:18} passes between the poles arising from the $\Gamma\left(b_j+s\right)$ and $\Gamma\left(1-a_j-s\right)$ gamma functions.  In our particular case, $\gamma$ encloses the poles at $s = -m+5/12,\, -m-1/12$, and $-m-1$ for $m = 0,\,1,\,\cdots$, but excludes the pole at $s = 0$.  In \eq{eq:17}, the contour $C$ includes this pole.  Therefore, to write the integral in \eq{eq:17} in terms of a Meijer G-function, we need to add the $s=0$ pole contribution to $G_{p,q}^{m,n}$.  Putting \eq{eq:17} into the form of \eq{eq:18} and applying Cauchy's integral formula~\cite{Arfken2013,Gbur_2011} reveals
\begin{widetext}
\begin{equation}\label{eq:19}
\begin{gathered}
\Langle T_{GC}^2 \Rangle = -2^{-5/6}\frac{55}{9\pi} \Gamma\begin{bmatrix}11/12\\-5/12,11/6,4/3\end{bmatrix}\sigma_\chi^2 \Langle T_G^2 \Rangle N_F^{5/6} \left\{\frac{9}{5} \sqrt{2\pi} \Gamma\begin{bmatrix}-5/6,13/6\\11/6\end{bmatrix} \right. \hfill \\
\left. \quad -\, G^{3,5}_{5,7}\left(N_F^2\left|\begin{matrix}1/6,-1/12,-7/12,1,-1/3\\-5/12,1/12,1,-5/6,-4/3,1/12,-5/12\end{matrix}\right.\right)\right\}, \hfill 
\end{gathered}    
\end{equation}
\end{widetext}
where $\sigma_\chi^2$ is the spherical-wave log-amplitude variance~\cite{Fried:67}.  Recall that $\Langle T_C^2\Rangle_\text{N}$ is the sum of the G-tilt variance $\Langle T_G^2 \Rangle$ in \eq{eq:9} with $\Langle T_{GC}^2 \Rangle$ in \eq{eq:19}.  

It is important to note that $\Langle T_C^2\Rangle_\text{N}$ is equivalent to the C-tilt variance integral expression derived by Holmes~\cite{Holmes:09}. He assumes that the denominator of \eq{eq:3} is equal to one.  This assumption is accurate in weak turbulence and provides a very good approximation for the C-tilt variance.  Here, we include the denominator because we are also interested in the G-tilt, C-tilt error, where its impact becomes important.   

\subsubsection{The denominator}
We now proceed to the denominator of \eq{eq:3}, the evaluation of which follows the same steps as above.  We make the weak turbulence approximation $\exp\left[ 4 B_{\chi}\left({\rho}\right) \right] \approx 1 + 4 B_{\chi}\left({\rho}\right)$, which splits the denominator into two integrals:
\begin{equation}
\begin{gathered}
\Langle T_C^2\Rangle_\text{D} \approx \frac{2\pi}{A} \int_0^\infty \rho \Lambda\left(\rho/D\right) \text{d}\rho \hfill \\
\quad +\, \frac{2\pi}{A} \int_0^\infty \rho \Lambda\left(\rho/D\right) 4B_\chi\left(\rho\right) \text{d}\rho. \hfill
\end{gathered}
\end{equation}
The first integral is unity, and the second is the aperture-averaged scintillation $\sigma_{\chi \text{A}}^2$~\cite{Ishimaru:99,Sasiela:07,Andrews:05}.

Focusing on $\sigma_{\chi \text{A}}^2$, we make the substitution $\rho = Dx$, multiply and divide $4 B_\chi\left(Dx\right)$ by $\left(Dx\right)^{-5/3}$, and apply the Mellin convolution theorem yielding (after simplification)
\begin{equation}\label{eq:25}
\begin{gathered}
\sigma_{\chi \text{A}}^2 = -\frac{2^{-5/2}}{\sqrt{\pi}} \frac{5}{9} \Gamma\begin{bmatrix}5/6\\2/3,11/3\end{bmatrix} C_n^2 k^2 z D^{5/3} \hfill \\
\;\times \frac{1}{\text{j}2\pi} \int_C \left(N_F^2\right)^{-s}  \Gamma\begin{bmatrix}-5/12+s,1/12+s,1+s\end{bmatrix} \hfill \\
\;\times \, \Gamma\begin{bmatrix}-s,7/6-s,4/3-s,5/3-s,11/6-s\\17/12-s,23/12-s,23/12-s,29/12-s\end{bmatrix} \text{d}s, \hfill
\end{gathered}
\end{equation}
where $C$ crosses the real $s$ axis between $5/12 < \operatorname{Re}\left(s\right) < 19/24$.  Like \eq{eq:19} above, to express \eq{eq:25} in the form of a Meijer G-function, we must add the $s=0$ residue to obtain the correct result.  Applying Cauchy's integral formula produces
\begin{widetext}
\begin{equation}\label{eq:26}
\begin{gathered}
\sigma_{\chi \text{A}}^2 = -\frac{2^{19/6}}{\pi^{3/2}} \Gamma\begin{bmatrix}11/12\\11/6,11/6,-5/12\end{bmatrix} \sigma_\chi^2 N_F^{5/6}\left\{ 2^{7/2}\sqrt{\pi}\Gamma\begin{bmatrix}-5/6,7/3,8/3\\17/6,23/6\end{bmatrix} \right. \hfill \\
\left. \quad -\, G^{3,5}_{5,7}\left(N_F^2\left|\begin{matrix}1,-1/6,-1/3,-2/3,-5/6\\-5/12,1/12,1,-5/12,-11/12,-11/12,-17/12\end{matrix}\right.\right)\right\}. \hfill
\end{gathered}   
\end{equation}
\end{widetext}
Finally, the C-tilt variance is
\begin{equation}\label{eq:27}
\Langle T_C^2 \Rangle = \frac{\Langle T_C^2\Rangle_\text{N}}{\Langle T_C^2\Rangle_\text{D}} = \frac{\Langle T_G^2\Rangle + \Langle T_{GC}^2\Rangle}{1+\sigma_{\chi\text{A}}^2},    
\end{equation}
where the G-tilt variance $\langle T_G^2 \rangle$, G-tilt, C-tilt variance $\langle T_{GC}^2\rangle$, and the aperture-averaged scintillation $\sigma_{\chi\text{A}}^2$ are in Eqs.~\eqref{eq:9},~\eqref{eq:19}, and~\eqref{eq:26}, respectively.  

\subsection{Asymptotic expressions for the C-tilt variance}
Although we have obtained a closed-form answer for the C-tilt variance, the result provides little insight into how $\Langle T_C^2\Rangle$ behaves versus $N_F$.  Inspection of the integrands in \eqs{eq:17}{eq:25} reveals that the integrals converge for all values of $N_F$ when the contours are closed to the left.  In both cases, the contours circumscribe the poles at $s = 0,\,-m+5/12,\, -m-1/12$, and $-m-1$ for $m = 0,\,1,\,\cdots$.

For $N_F \ll 1$ (i.e., the far-zone) the numerator of $\Langle T_C^2\Rangle$ is well approximated by summing the pole residues at $s = 0$, $5/12$, and $-1/12$.  Performing the requisite analysis yields
\begin{equation}\label{eq:31}
\begin{gathered}
\Langle T_{C}^2 \Rangle_\text{N} \approx \Langle T_G^2\Rangle \left\{1 + 4 \sigma_\chi^2 \right. \hfill \\
\left. \; +\, \frac{2^{5/3}}{\sqrt{\pi}}\frac{99}{55}\Gamma\begin{bmatrix}13/6,11/12,1/6\\5/6,11/6,4/3,-5/12\end{bmatrix}\sigma_\chi^2 N_F^{5/6} \right. \hfill \\
\left. \; -\, \frac{2^{7/6}}{{\pi}}\frac{3}{17}\Gamma\begin{bmatrix}1/6,11/3,1/12,11/12,11/12\\11/6,-5/12,4/3\end{bmatrix}\sigma_\chi^2 N_F\right\}. \hfill 
\end{gathered}
\end{equation}
Likewise, $\Langle T_C^2\Rangle_\text{D}$ is well approximated by summing the residues of the same three poles, namely,
\begin{equation}\label{eq:32}
\begin{gathered}
\Langle T_C^2\Rangle_\text{D} \approx 1 + 4\sigma_\chi^2 \hfill \\
\; -\, \frac{2^{13/2}}{\pi^2} \Gamma\begin{bmatrix}11/12,1/12,4/3,7/3\\11/6,17/6,23/6\end{bmatrix} \sigma_\chi^2 N_F^{5/6} \hfill \\
\; +\, \frac{2^{11/6}}{\pi} \Gamma\begin{bmatrix}1/12,23/12,-1/12,11/12\\11/6,-5/12\end{bmatrix} \sigma_\chi^2 N_F. \hfill
\end{gathered}    
\end{equation}
Nevertheless, dividing \eq{eq:31} by \eq{eq:32} yields a poor estimate for $\Langle T_C^2 \Rangle$.  Consequently, we use $\Langle T_C^2\Rangle_\text{D} \approx 1 + 4\sigma_\chi^2$ when we approximate $\Langle T_C^2 \Rangle$ for $N_F \ll 1$.  

For $N_F \gg 1$ (i.e., the near-zone) the sums over the pole residues are slow to converge and can result in numerical errors when evaluating the Meijer G-functions in \eqs{eq:19}{eq:26}.  We, therefore, seek asymptotic solutions to the integrals in \eqs{eq:17}{eq:25} that are accurate in the limit $N_F \to \infty$.  We obtain such solutions by including contributions from poles to the right of the contours.  

In \eq{eq:17}, the dominant contribution comes from the pole at $s=5/6$ and, in fact, provides an excellent approximation for $\Langle T_C^2\Rangle_\text{N}$ when $N_F > 1$.  Applying Cauchy's integral formula to compute the pole residue at $s=5/6$ yields the following expression for $\Langle T_C^2\Rangle_\text{N}$:        
\begin{equation}\label{eq:20}
\begin{gathered}
\Langle T_{C}^2 \Rangle_\text{N} \approx \Langle T_G^2\Rangle \left\{1 + \frac{40}{9\sqrt{\pi}} \right. \hfill \\
\left. \quad \times\, \Gamma\begin{bmatrix}1/12,5/6,11/6,17/6\\17/12,11/3,1/6\end{bmatrix} \sigma_\chi^2 N_F^{-5/6}\right\}.
\end{gathered}
\end{equation}
For $\Langle T_C^2\Rangle_\text{D}$, the dominant contribution comes from the pole at $s = 1$.  Applying Cauchy's integral formula, we obtain
\begin{equation}\label{eq:T31}
\Langle T_C^2\Rangle_\text{D} \approx 1 + \frac{8}{{\pi}^2} \Gamma\begin{bmatrix}13/12,1/6,1/3,2/3\\5/12,11/6,11/6\end{bmatrix} \sigma_\chi^2 N_F^{-7/6}. 
\end{equation}

As we can see in \eqs{eq:32}{eq:T31}, $\Langle T_C^2 \Rangle_\text{D}$ is slightly greater than one.  Indeed, assuming $\Langle T_C^2 \Rangle_\text{D} \approx 1$ yields a good, yet overestimated value for the C-tilt variance.  

\subsection{G-tilt, C-tilt error}
The G-tilt, C-tilt error $E_{GC}$ is a measure of the G-tilt, C-tilt anisoplanatism~\cite{Holmes:09,Mitchell:25}, which is an important metric for adaptive-optics system performance~\cite{Barchers:02,Banet:20,Kalensky:24}. We define this error as the root-mean-square difference between the G-tilt and C-tilt angles, such that
\begin{equation}\label{eq:28}
E_{GC} = \sqrt{\Langle \left| \boldsymbol{T}_G - \boldsymbol{T}_C \right|^2 \Rangle}.    
\end{equation}
It is easy to show using \eq{eq:T4} that $E_{GC}$ is equivalent to
\begin{equation}\label{eq:29}
E_{GC} = \sqrt{\Langle {T}_C^2 \Rangle - \Langle {T}_G^2 \Rangle}.    
\end{equation}
Note that the G-tilt angle $\boldsymbol{T}_G$ equals \eq{eq:T4} with $\exp\left[2\chi\left(\boldsymbol{\rho}\right)\right] = 1$.  Substituting \eq{eq:27} into \eq{eq:29} and simplifying yields
\begin{equation}\label{eq:T34}
E_{GC} = \sqrt{\frac{\Langle T_{GC}^2\Rangle - \sigma_{\chi\text{A}}^2 \Langle {T}_G^2 \Rangle}{1+\sigma_{\chi\text{A}}^2}}.
\end{equation}

Deriving approximate relations for the G-tilt, C-tilt error using the asymptotic expressions above is somewhat ad hoc because the value of the numerator is approximately zero.  Nevertheless, we can obtain relatively accurate relations for $E_{GC}$.  

Staring with $N_F \ll 1$, substituting the relevant parts of \eqs{eq:31}{eq:32} into the numerator and keeping only the first two terms yields
\begin{equation}\label{eq:T35}
E_{GC} \approx 1.4017 N_F^{5/12} \sqrt{\frac{\sigma_\chi^2 \Langle T_G^2\Rangle}{\Langle T_C^2\Rangle_\text{D}}} \quad N_F \ll 1,
\end{equation}
where we use all of \eq{eq:32} in the denominator of \eq{eq:T35}.

For $N_F \gg 1$, we substitute the germane parts of \eqs{eq:20}{eq:T31} into the numerator of \eq{eq:T34} and keep only the $N_F^{-5/6}$ term.  After simplification, the result is
\begin{equation}\label{eq:T36}
E_{GC} \approx 1.6331 N_F^{-5/12} \sqrt{\frac{\sigma_\chi^2 \Langle T_G^2\Rangle}{\Langle T_C^2\Rangle_\text{D}}} \quad N_F \gg 1,    
\end{equation}
where we use all of \eq{eq:T31} in the denominator.

\section{Simulations}\label{sec:sim}
To validate the above theory, we employed wave-optics simulations. In practice, we simulated the propagation of a $\lambda = 1 \text{ } \mu\text{m}$ point source $z = 10 \text{ km}$ through atmospheric turbulence.  Fried's parameter $r_0$~\cite{Fried1966} and $\sigma_\chi^2$ were $9.04 \text{ cm}$ and $0.2$, respectively.  We changed the pupil diameter $D$ to vary the Fresnel number $N_F$ from $0.5$ to $50$ in 9 logarithmically spaced steps.  

We used the split-step method~\cite{Schmidt:10} to simulate wave propagation for each $N_F$.  To avoid aliasing, the simulated fields required grids that were $850$ points on a side.  The grid spacings and number of partial propagations depended on $N_F$ and are given in Table~\ref{tab:sim_params}.

\begin{table}
    \centering
    \caption{Simulation Parameters: Fresnel Number $N_F$, Pupil/Aperture Diameter $D$, Source-Plane Grid Spacing $\delta_\text{src}$, Pupil-Plane Grid Spacing $\delta_\text{pup}$, Number of Partial Propagations $n$}
    \begin{tabular}{|c|c|c|c|c|}
    \hline
     $N_F$ & $D \text{ (cm)}$ & $\delta_\text{src} \text{ (mm)}$ & $\delta_\text{pup} \text{ (mm)}$ & n \\
       \hline
      0.500 & 7.98 & 10.4 & 0.997 & 13 \\
      0.889 & 10.6 & 7.83 & 1.33 & 8 \\
      1.58  & 14.2 & 5.87 & 1.42 & 7 \\
      2.81  & 18.9 & 4.40 & 1.89 & 5 \\
      5.00  & 25.2 & 3.30 & 2.52 & 4 \\
      8.89  & 33.6 & 2.48 & 3.36 & 4 \\
      15.8  & 44.9 & 1.86 & 4.49 & 5\\
      28.1  & 59.8 & 1.39 & 5.98 & 8\\
      50.0  & 79.8 & 1.04 & 7.98 & 12\\
       \hline
    \end{tabular}    
    \label{tab:sim_params}
\end{table}

We determined the strengths or $r_0$'s of the phase screens using the linear optimization method described in Ref.~\cite{Schmidt:10}.  The number of phase screens used in a particular $N_F$ simulation was equal to the number of partial propagations given in Table~\ref{tab:sim_params}.  We generated each screen using the Fourier/spectral method augmented with subharmonics to account for low-spatial-frequency aberrations, especially tilt~\cite{doi:10.1088/0959-7174/2/3/003,Frehlich:00,Carbillet:10,Charnotskii:20}.  We used the modified von K\'{a}rm\'{a}n~\cite{Andrews:05,Sasiela:07} index of refraction power spectrum to synthesize the phase screens, where the inner and outer scales were unphysically small and large ($l_0 = \lambda/2$ and $L_0 = 500 \text{ km}$), respectively.   The purpose of this was to closely approximate the Kolmogorov spectrum, which we used to derive the theoretical expressions in Section~\ref{sec:thy}.

After collimating the field $U$ in the pupil plane, we windowed $U$ using a circle function with a diameter $D$ equal to that of the corresponding $N_F$.  Then, we computed both the G-tilt and C-tilt using Lukin's expression~\cite{Lukin}
\begin{equation}\label{eq:T37}
\begin{gathered}
    \boldsymbol{T} = \frac{1}{k}\frac{\displaystyle \iint_A \operatorname{Re}\left[U\left(\boldsymbol{\rho}\right)\right]\nabla \operatorname{Im}\left[U\left(\boldsymbol{\rho}\right)\right] \text{d}^2\rho}{\displaystyle \iint_A \left|U\left(\boldsymbol{\rho}\right)\right|^2 \text{d}^2\rho} \hfill \\
\; -\, \frac{1}{k} \frac{\displaystyle\iint_A\operatorname{Im}\left[U\left(\boldsymbol{\rho}\right)\right]\nabla\operatorname{Re}\left[U\left(\boldsymbol{\rho}\right)\right]\text{d}^2\rho}{\displaystyle \iint_A \left|U\left(\boldsymbol{\rho}\right)\right|^2 \text{d}^2\rho}. \hfill
\end{gathered}
\end{equation}
In \eq{eq:T37}, we computed the gradients using fast Fourier transforms and evaluated the integrals over the pupil area $A$ using Simpson's rule.  We computed the C-tilt by directly applying \eq{eq:T37}.  For the G-tilt, we extracted the complex argument of $U$, ``flattened'' its amplitude, and then applied \eq{eq:T37}.

We computed both the G-tilt and C-tilt statistics from $10,000$ independent turbulence realizations (trials) at each $N_F$.  We then divided the $10,000$ trials into partitions of $500$, computed the G-tilt and C-tilt variances over each partition, and averaged the variances.  We also computed the standard deviation of the $20$ G-tilt and C-tilt variances to show the $\pm 1\sigma$ bounds on those results.

\section{Results}\label{sec:res}
\begin{figure}
    \centering
    \includegraphics*[width=\columnwidth]{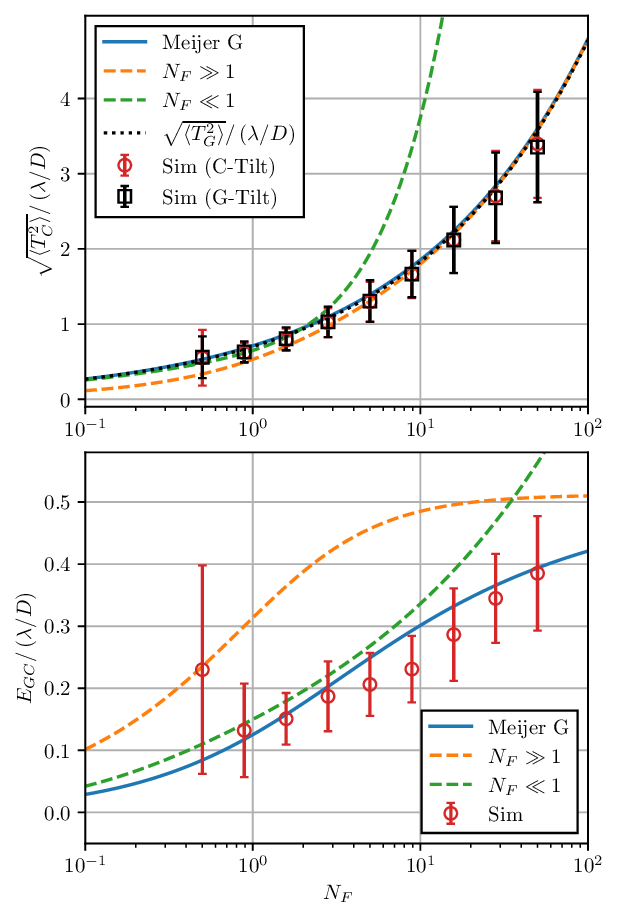}
    \caption{Normalized C-tilt standard deviation $\sqrt{\Langle T_C^2\Rangle}/\left(\lambda/D\right)$ (top) and normalized G-tilt, C-tilt error $E_{GC}/\left(\lambda/D\right)$ (bottom) results.  The bars are $\pm 1 \sigma$.  }
    \label{fig:sim}
\end{figure}

Figure~\ref{fig:sim} shows the results.  In the top plot, we display the C-tilt standard deviation (also known as the noise-equivalent angle \cite{Mitchell:25}) normalized by the diffraction-limited angular beam width $\lambda/D$.  The ``Meijer G,'' ``$N_F \gg 1$,'' and ``$N_F \ll 1$'' traces are theoretical predictions using Eqs.~\eqref{eq:27} [in combination with Eqs.~\eqref{eq:9},~\eqref{eq:19}, and~\eqref{eq:26}], \eqs{eq:20}{eq:T31}, and \eqs{eq:31}{eq:32}, respectively.  We have also included the normalized G-tilt angle standard deviation as the black dotted curve [see \eq{eq:9}].  Lastly, the red circles and black squares, both with error bars, are the C-tilt and G-tilt simulation results.

The bottom plot shows the normalized G-tilt, C-tilt error versus $N_F$.  The ``Meijer G,'' ``$N_F \gg 1$,'' and ``$N_F \ll 1$'' traces are theoretical predictions using \eq{eq:T34} [again, in combination with Eqs.~\eqref{eq:9},~\eqref{eq:19}, and~\eqref{eq:26}], \eq{eq:T36}, and \eq{eq:T35}.  The red circles with error bars are the simulated results computed using the raw G-tilt and C-tilt angles and \eq{eq:28}.

Overall, the agreement between the simulated and theoretical results is quite good.  We can explain the minor differences between the curves by a combination of the approximations used to derive the theoretical expressions and the well-documented difficulty in properly including low-spatial-frequency aberrations in atmospheric turbulence simulations~\cite{doi:10.1088/0959-7174/2/3/003,Carbillet:10,Charnotskii:20}.

Before concluding the paper, we would like to point out that while both the normalized and unnormalized G-tilt, C-tilt error approach zero as $N_F \to 0$, the near-zone limits of the two $E_{GC}$ are different.  As $N_F \to \infty$, the unnormalized $E_{GC}$, like its corresponding far-zone expression, approaches zero.  However, in the normalized case, the $1/N_F$ decrease of $E_{GC}$ is exactly canceled by the $1/N_F$ decrease of $\lambda/D$ yielding the asymptotic value
\begin{equation}
\lim\limits_{N_F \to \infty}\frac{E_{GC}}{\lambda/D} = \frac{2^{10/3}}{\pi^{7/4}}\sigma_\chi^2\sqrt{\Gamma\begin{bmatrix}1/12,4/3,11/12\\11/6,5/12,7/12\end{bmatrix}} .    
\end{equation}
In physical units (radians), the $E_{GC}$ initially increases with $N_F$, reaches a maximum near $N_F = 1$, and then decreases toward zero for $N_F > 1$.  

\section{Conclusion} \label{sec:con}
In this paper, we derived three closed-form expressions for the spherical-wave C-tilt variance $\Langle T_C^2\Rangle$ in constant-$C_n^2$ Kolomogorov atmospheric turbulence.  All three expressions were functions of the Fresnel number $N_F$.  The first was applicable for all $N_F$ and was in the form of a Meijer G-function.  The second and third were much simpler relations and valid in the asymptotic limits $N_F \to 0$ and $N_F \to \infty$ (i.e., the far- and near-zones, respectively).  In addition, we investigated the difference between G-tilt and C-tilt and derived corresponding relations for the G-tilt, C-tilt error $E_{GC}$ due to a scintillated spherical wave.  Lastly, we validated our analysis by comparing the theory with wave-optics simulations. Our $\Langle T_C^2 \Rangle$ and $E_{GC}$ predictions were in very good agreement with the simulation results.  

The rigorous analysis of C-tilt presented in this paper will find use in quantifying the performance of centroid-based trackers and wavefront sensors, which are critical components in adaptive-optics systems.

\begin{acknowledgments}
The authors would like to thank the Joint Directed Energy Transition Office for sponsoring this research. Approved for public release; distribution is unlimited. Public Affairs release approval SMDC PAO$\#$5063. 
M.W.H.: The views expressed in this paper are those of the author and do not reflect the policy or position of Epsilon C5I or Epsilon Systems. 
\end{acknowledgments}

\bibliography{main_APS}
\end{document}